\pdfoutput=1

\documentclass[11pt, reqno,preprint]{article}
\usepackage{jheppub}
\usepackage{epsfig}
\usepackage{amssymb}
\usepackage{amsmath}
\usepackage{mathrsfs}
\usepackage{hyperref}
\usepackage{multirow}
\usepackage{dsfont}

\def\be{\begin{equation}}
\def\ee{\end{equation}}
\def\ba{\begin{eqnarray}}
\def\ea{\end{eqnarray}}
\def\nl{\nonumber\\}

\def\a{\alpha}
\def\b{\beta}

\def\l{\langle}
\def\r{\rangle}

\def\b#1{\overline{#1}}

\def\CP1{\mathbb{CP}^1}
\def\SL2C{\mathrm{SL}(2,\mathbb{C})}
\def\GLoneC{\mathrm{GL}(1,\mathbb{C})}
\def\GLtwoC{\mathrm{GL}(2,\mathbb{C})}
\def\Z2{\mathbb{Z}_2}
\def \bangle{ \atopwithdelims \langle \rangle}

\title{Scattering in Three Dimensions from Rational Maps}
\author{Freddy Cachazo${}^{a}$, Song He${}^{a,b}$ and Ellis Ye Yuan${}^{a,c}$}
\affiliation[a]{Perimeter Institute for Theoretical Physics, Waterloo, ON N2L 2Y5,
Canada}
\affiliation[b]{School of Natural Sciences,
Institute for Advanced Study, Princeton, NJ 08540, USA}
\affiliation[c]{Physics Department, University of Waterloo, Waterloo, ON N2L 2Y5,
Canada}
\emailAdd{fcachazo, she, yyuan@perimeterinstitute.ca}

\abstract{The complete tree-level S-matrix of four dimensional ${\cal N}=4$ super Yang-Mills and ${\cal N} = 8$ supergravity has compact forms as integrals over the moduli space of certain rational maps.
In this note we derive formulas for amplitudes in three dimensions by using the fact that when amplitudes are dressed with proper wave functions dimensional reduction becomes straightforward. This procedure leads to formulas in terms of rational maps for three dimensional maximally supersymmetric Yang-Mills and gravity theories. The integrand of the new formulas contains three basic structures: Parke-Taylor-like factors, Vandermonde determinants and resultants. Integrating out some of the Grassmann directions produces formulas for theories with less than maximal supersymmetry, which exposes yet a fourth kind of structure. Combining all four basic structures we start a search for consistent S-matrices in three dimensions. Very nicely, the most natural ones are those corresponding to ABJM and BLG theories. We also make a connection between the power of a resultant in the integrand, representations of the Poincar\'e group, infrared behavior and conformality of a theory. Extensions to other theories in three dimensions and to arbitrary dimensions are also discussed.}

\begin{document}
\maketitle

\section{Introduction: Scattering in Four Dimensions}

The complete tree-level S-matrix of maximally supersymmetric Yang-Mills and gravity in four dimensions can be presented as integrals over the moduli space of maps from the $n$-punctured sphere to twistor space~\cite{Witten:2003nn,Roiban:2004yf,Cachazo:2012kg} or to momentum space~\cite{Cachazo:2012da,Cachazo:2013zc}. In this note we consider the version that maps into momentum space when scattering amplitudes are dressed with wave functions imposing on-shell conditions on the scattering data. With the new form, dimensional reduction down to three dimensions can be done straightforwardly by integrating out the extra momentum dimension as explained in section 2. Applying this technique to the maximally supersymmetric theories, we find the corresponding theories in three dimensions. In the case of maximal supergravity we find that our reduction procedure directly forces the amplitude to vanish whenever it involves odd number of particles or whenever the R-charge sector of the parent theory is not the helicity conserving one.

Furthermore, theories with less amount of supersymmetries can be obtained by integrating out Grassmann directions. This leads to the appearance of a new structure depending on the location of the punctures that is not present in four dimensions, which motivate us to start out a search for consistent theories in three dimensions. In this context, consistency means correct factorization properties. Very nicely, the Aharony-Bergman-Jafferis-Maldacena (ABJM)~\cite{Aharony2008a} and Bagger-Lambert-Gustavsson (BLG) theories~\cite{Bagger2007a,Gustavsson2007a} come about in a natural way as discussed in section 3. We also prove that our formula for ABJM is equivalent to that obtained from an integral over regions of $\mathrm{G(2,n)}$ which map to an orthogonal Grassmannian~\cite{Huang:2012vt}. 

In order to present the new form of the four dimensional amplitudes and to establish notations, let us write the on-shell momentum space as
\be \{k^\mu_1,\ldots,k^\mu_n ~ | ~ \sum^n_{a=1} k^\mu_a=0,~
k^2_1=\ldots=k^2_n=0\}
\ee
which implies that each $k_{\underline{\alpha}\underline{\dot\alpha}}=k_{\mu}\sigma^{\mu}_ {\underline{\alpha}\underline{\dot\alpha}}= \lambda_{\underline{\alpha}}\tilde\lambda_{\underline{\dot \alpha}}$, with $\underline{\alpha},\underline{\dot \alpha}\in \{1,2\}$ denoting Lorentz spinor indices. The on-shell super-space is then
\be \{\theta^I_{1,\underline{\alpha}},\tilde\theta^I_{1,\underline{\dot\alpha}},\ldots, \theta^I_{n,\underline{\alpha}},\tilde\theta^I_{n,\underline{\dot\alpha}} ~ | ~
\sum^n_{a=1}\theta^I_{a,\underline{\alpha}}=\sum^n_{a=1}\tilde\theta^I_{a,\underline{\dot\alpha}}=0, ~ \l\lambda_a \theta^I_a\r=[\tilde\lambda_a\tilde\theta^I_a]=0\}\ee
with $I = 1,\ldots ,{\cal N}/2$ and the amount of supersymmetries ${\cal N}$ is related to the maximum helicity in the theory by ${\cal N}= 4s$. Here we have chosen a parity invariant formulation which only makes manifest a subgroup $SU({\cal N}/2)\times SU({\cal N}/2)$ of the $SU({\cal N})$ R-symmetry group. The on-shell conditions imply that one can write $\theta_{a,\underline{\alpha}}^I = \lambda_{a,\underline{\alpha}}\tilde\eta^I_a$ and $\tilde\theta_{a,\underline{\dot\alpha}}^I = \tilde\lambda_{a,\underline{\dot\alpha}}\eta^I_a$.

If $M_{n,k}$ denotes the standard $n$-particle amplitude in the $k$ R-charge sector then our object of study is
\be
{\cal M}_{n,k} = \prod_{a=1}^n \delta(k^2_a)\delta^{0|\frac{\mathcal{N}}2}(\l\lambda_a
\theta_a^I\r)\delta^{0|\frac{\mathcal{N}}2}([\tilde\lambda_a\tilde\theta_a^I])M_{n,k}. \label{wavefunction}
\ee
Here $M_{n,k}$ is assumed to be a distribution with support on momentum conserving external momenta.

Let $\mathbb{L}(z)$ and $\tilde{\mathbb{L}}(z)$ be rational maps from $\mathbb{CP}^1$ to $\mathbb{CP}^{1|{\cal N}/2}$ of degree $d=k-1$ and $\tilde d=n-k-1$ respectively. More explicitly, $\mathbb{L}(z) = (\lambda_{\underline{\alpha}}(z),\eta^I (z))$ with
\be
\lambda_{\underline{\alpha}}(z) = \sum_{\alpha=0}^d \rho_{\alpha,\underline{\alpha}}z^\alpha, \quad \eta^I(z) = \sum_{\alpha=0}^d \chi^I_{\alpha}z^\alpha~\label{map},
\ee
and similar expressions hold for the tilde variables. In order to simplify the presentation of the formulas it is convenient to introduce $\mathbb{M}_\alpha \equiv(\rho_\alpha|\chi_\alpha )$ and $\tilde{\mathbb{M}}_\alpha \equiv (\tilde\rho_\alpha |\tilde\chi_\alpha )$ to represent the collection of coefficients of $z^\alpha$ in the maps.

Finally, the dressed scattering amplitude ${\cal M}^{(s)}_{n,k}$ can be presented as
\be
\!\!\int\!\! \frac{d^n\sigma\,d
\mathbb{M}~d\tilde{\mathbb{M}}}{\textrm{vol }G}\prod^n_{a=1}\delta^4(k_a-\oint_a dz
\frac{k(z)}{P_n(z)})\delta^{0|\mathcal{N}}(\theta_a-
\oint_a dz\frac{\theta(z)}{P_n(z)})\delta^{0|\mathcal{N}}(\tilde\theta_a-\oint_a dz
\frac{\tilde\theta(z)}{P_n(z)})I_s~\label{4dformula}
\ee
with
$$k_{\underline{\alpha},\underline{\dot\alpha}}(z) = \lambda_{\underline{\alpha}}(z)\tilde\lambda_{\underline{\dot\alpha}}(z),\quad \theta^I_{\underline{\alpha}}(z) = \lambda_{\underline{\alpha}}(z)\tilde\eta^I(z),\quad  \tilde\theta_I^{\underline{\dot\alpha}}(z) = \tilde\lambda^{\underline{\dot\alpha}}(z)\eta_I(z),$$
and where
\be
d\mathbb{M} = \prod_{\alpha=0}^d d^{2|\frac{\cal N}{2}}\mathbb{M}_\alpha, \quad d\tilde{\mathbb{M}} = \prod_{\alpha=0}^{\tilde d} d^{2|\frac{\cal N}{2}}\tilde{\mathbb{M}}_\alpha,\quad \oint_a dz = \frac{1}{2\pi i}\int_{|z-\sigma_a|=\epsilon}\!\!\! dz, \quad P_n(z) = \prod_{b=1}^n(z-\sigma_b).
\ee
Finally, there is a gauge redundancy given by the group
\be
G=\SL2C\times \GLoneC.
\ee
The integrand $I_s$ depends on the theory and we devote the next subsection to a short discussion of the properties of Yang-Mills and gravity theories that lead to its definition. Before turning to the integrand let us comment on how~(\ref{4dformula}) can be used to compute ${\cal M}^{(s)}_{n,k}$ in practice. There are $4n$ bosonic equations
\be
k_a^{\underline{\alpha},\underline{\dot\alpha}} = \oint_a dz \frac{\lambda^{\underline{\alpha}}(z)\tilde\lambda^{\underline{\dot\alpha}}(z)}{\prod_{b=1}^n(z-\sigma_b)}.
\ee
Of these equations, $n+4$ constrain the external momenta to be on-shell, i.e.~$k^2_a=0$, and momentum conserving. The remaining $3n-4$ constrain the integration variables. The positions of the punctures, $\sigma_a$, modulo $\SL2C$ give $n-3$ while the coefficients $\rho$ and $\tilde\rho$ give $2(d+1)$ and $2({\tilde d}+1)$ respectively. Using that $d+{\tilde d}=n-2$ and that there is a $\GLoneC$ redundancy $\{\lambda(z),\tilde\lambda(z)\}\to\{ t \lambda(z),t^{-1}\tilde\lambda(z)\}$ one gets exactly $3n-4$ variables. This means that the amplitude is the summation of evaluations of a Jacobian on all solutions to the system of equations
\be  \lambda^{\underline{\alpha}}(\sigma_a)\tilde\lambda^{\underline{\dot\alpha}}(\sigma_a)=k_{a}^{\underline{\alpha}\underline{\dot\alpha}}\prod_{b\neq a} (\sigma_a-\sigma_b), \quad \textrm{for}~a=1,\ldots, n. ~\label{4deq}\ee

\subsection{The Integrand $I_s$ and Physical Constraints}~\label{sec:4dintegrands}

In general, very strong constraints are imposed on $I_s$ by the fact that the full integral ${\cal M}_{n,k}$ should be $\SL2C$ invariant and must have simple poles together with correct residues on factorization singular limits. Let us start with supergravity. In this case there is a direct hint on what kind of structure should be present. Note that the maps define a $z$-dependent vector $k_{\underline{\alpha},\underline{\dot\alpha}}(z) = \lambda_{\underline{\alpha}}(z)\tilde\lambda_{\underline{\dot\alpha}}(z)$ which can be thought of as mapping $\mathbb{CP}^1$ into the null cone in the momentum space. Clearly, there exists vectors $k_{\underline{\alpha},\underline{\dot\alpha}}(z)$ that map certain points in $\mathbb{CP}^1$ to the tip of the cone, i.e., to the zero momentum vector. Recall that in the representation theory of the Poincar\'e algebra a non-vanishing null vector has as little group\footnote{Here we ignore the translation generators as they are trivially represented.} $SO(2)$ while the zero vector has $SO(3,1)$. This means that they induce different representations and for graviton amplitudes we should remove this zero vector from the integration region. Clearly, the only way  $k_{\underline{\alpha},\underline{\dot\alpha}}(z)$ can map a point in $\mathbb{CP}^1$ to the zero vector is if both components of either $\lambda_{\underline{\alpha}}(z)$ or $\tilde\lambda_{\underline{\dot\alpha}}(z)$ can vanish simultaneously, or equivalently the two component polynomials of $\lambda_{\underline{\alpha}}(z)$ or $\tilde\lambda_{\underline{\dot\alpha}}(z)$ share at least one common root. This condition is nothing but the vanishing of the resultant of the two polynomials. And therefore we conclude that for supergravity theories we must have
\be I_{s=2}=R(\lambda_{\underline{1}}(z),\lambda_{\underline{2}}(z),z)
R(\tilde\lambda_{\underline{\dot{1}}}(z),\tilde\lambda_{\underline{\dot{2}}}(z),z)~\label{resultants} \ee
where $R(f(z),g(z),z)$ denotes the resultant of $f(z)$ and $g(z)$ with respect\footnote{A familiar example is the resultant of a polynomial $f(z)$ and its derivative $f'(z)$ which gives the discriminant of $f(z)$.} to $z$ (see appendix \ref{sec:resultant} for a general definition and some examples of resultants). As it turns out, no other factors are needed in $I_2$ to ensure $\SL2C$ invariance and proper factorizations of the amplitude. So the most natural choice of $I_2$ is also the correct one.

Turning to super-Yang-Mills we note that the theory is conformal i.e., invariant under the larger $SO(4,2)$ group. This means that the zero momentum vector is not a special point anymore and does not have to be excluded. Therefore no resultants should appear as factors in $I_1$. A different constraint comes from the fact that the amplitude can be decomposed in partial amplitudes by stripping out color factors and each partial amplitude is only invariant under cyclic permutations of the labels with a given order. Take for example the canonical order $(1,2,\cdots ,n-1,n)$. Everything in the definition of ${\cal M}^{(1)}_{n,k}$ so far is completely permutation invariant. This means that $I_1$ must break the full permutation invariance down to only the cyclic invariance. The simplest choice of the integrand that leads to a $\SL2C$ invariant formula is
\be I_{s=1}=\frac 1{(\sigma_{1}-\sigma_2)(\sigma_2-\sigma_3)\ldots (\sigma_{n}-\sigma_1)} \label{parke}\ee
and just as in gravity it turns out to be the correct choice.

\subsection{Counting the Number of Solutions}

Before going to three dimensions, let us count the number of solutions of (\ref{4deq}), or equivalently the number of terms when we compute amplitudes using (\ref{4dformula}). In~\cite{Spradlin:2009qr} it was conjectured that the number of solutions, for amplitudes with $n$ points and R-charge sector $k=d{+}1$, is the Eulerian number $N_{d,\tilde d}={d{+}\tilde d{-}1 \bangle d{-}1}={n{-}3 \bangle k{-}2}$. Recall $\tilde d=n{-}d{-}2$, and as a consequence, the total number of solutions for $n$ points in all the sectors is $\sum^{n{-}2}_{k=2} {n{-}3 \bangle k{-}2}=(n{-}3)!$ (For more properties of Eulerian numbers see ref.~\cite{EulerianNumbers}). Here we prove the conjecture inductively by studying soft limits of (\ref{4deq}) where $\lambda(z)$ and $\tilde\lambda(z)$ have degree $d$ and $\tilde d$ respectively.

In the soft limit $k_n\to 0$, the last equation implies either $\lambda(\sigma_n)=0$ or $\tilde\lambda(\sigma_n)=0$, and we first consider the holomorphic limit. $\lambda^{\underline1}(\sigma_n)=\lambda^{\underline2}(\sigma_n)=0$ means the two components share a common root $\sigma_n$, thus $\lambda(z)=(z-\sigma_n)\lambda^*(z)$ for some degree-$(d{-}1)$ polynomial spinor $\lambda^*(z)$.  In the remaining $n{-}1$ equations, the factor $\sigma_a-\sigma_n$ can be removed from both sides and particle $n$ completely decouples from the equations
\be\label{eq:softlimiteqns} \lambda^{*\underline{\alpha}}(\sigma_a)\tilde\lambda^{\underline{\dot\alpha}}(\sigma_a)=k_{a}^{\underline{\alpha}\underline{\dot\alpha}}\prod_{b\neq a,n}(\sigma_a-\sigma_b),\ee for  $a=1,\ldots, n-1$. Note that these are exactly the system of equations with $n{-}1$ points and bi-degree $(d{-}1,\tilde d)$. In other words, we have taken a $k$-decreasing soft limit, and by our induction assumption the number of solutions for the $(n{-}1)$-point equations is $N_{d{-}1,\tilde d}$.

Now we study the last equation near the holomorphic soft limit; writing $k_n^{\underline{\alpha}\underline{\dot{\alpha}}}=\lambda^{\underline{\alpha}}_n \tilde\lambda^{\underline{\dot{\alpha}}}_n$, we keep $\tilde\lambda_n$ finite and take $\lambda_n=\epsilon \xi_n \to 0$ as $\epsilon\to 0$. It is easy to see that $\lambda(\sigma_n)\propto \epsilon \xi_ n$ will drop out from the last equation. By taking e.g. the ratio of components $\lambda_n^{\underline 1 }\tilde\lambda_n^{\underline{\dot 1}}$ and $\lambda_n^{\underline 1}\tilde\lambda_n^{\underline{\dot 2}}$, the equation becomes $[\tilde\lambda(\sigma_n),\tilde\lambda_n]=0$. Up to higher-order terms in $\epsilon$, we can plug in a solution $(\rho, \tilde\rho, \sigma_1,\ldots,\sigma_{n{-}1})^{(i)}$ ($i=1,\ldots, N_{d{-}1,\tilde d}$) from solving $n{-}1$ equations into the last one, and solve
\be [\tilde \lambda^{(i)}(\sigma_n), \tilde\lambda_n]=0,\ee which is a degree-$\tilde d$ polynomial equation for $\sigma_n$, and there are $\tilde d$ solutions. In total, we have $\tilde d\times N_{d{-}1,\tilde d}$ solutions near the $k$-decreasing soft limit. Similarly the anti-holomorphic case gives the $k$-preserving soft limit, where we have $d\times N_{d,\tilde d{-}1}$ solutions since the last equation becomes a degree-$d$ polynomial equation for $\sigma_n$. Since the number of solutions cannot change abruptly as we move away from the soft limit, we conclude that for generic external momenta the number of solutions satisfies the recursion
\be N_{d,\tilde d}=\tilde d \times N_{d{-}1,\tilde d}+ d \times N_{d,\tilde d{-}1}, \ee which together with the initial conditions $N_{1,1}=N_{1,2}=N_{2,1}=1$, is exactly the recursion relation for Eulerian numbers. Obviously $N_{d,\tilde d}=N_{\tilde d, d}$, $N_{1,n{-}3}=1$, and a few more examples are $N_{2,2}=4$ for $n=6$, $N_{2,3}=11$ for $n=7$, and $N_{2,4}=26, N_{3,3}=66$ for $n=8$.

\section{Dimensional Reduction}

The main advantage of working with amplitudes dressed by wave functions is that dimensional reduction is straightforward. The general procedure is to start from the four-dimensional momentum $k^{\underline{\alpha}\dot{\underline\alpha}}$,
integrate out the component $k^{\underline 2 \dot {\underline 1}}$ (or alternatively $k^{\underline 1 \dot {\underline 2}}$) with the
constraint $\delta(k^{\underline 1 \dot {\underline 2}}-k^{\underline 2 \dot {\underline 1}})$, and arrive at the three-dimensional momentum
$k^{\underline{\alpha} \underline{\beta}}$ defined as $\{k^{\underline 1\underline 1},k^{\underline 1\underline 2}=k^{\underline 2\underline
1},k^{\underline 2\underline 2}\}$\,.
This should be done for, e.g. the first $n-1$ particles, and for particle $n$ we simply do the integration without imposing more constraint, since the momentum conservation in four dimensions already implies that $k^{\underline 1 \dot {\underline 2}}_n=k^{\underline 2 \dot {\underline 1}}_n$, given this is true for the other particles. And so one component of the momentum conservation is used up in the dimensional reduction, which is desired.
To be explicit, the integration over the $\underline {2}\dot {\underline 1}$-component of all $n$ particles
removes the unwanted delta function and gives,
\be \int \prod^n_{a=1} d k_a^{\underline{2} \underline{\dot 1}}
\prod^{n{-}1}_{b=1}\delta(k_b^{\underline{1}\underline{\dot2}}-k_b^{\underline{2}\underline{\dot1}}) {\cal M}^{D=4}_n={\cal M}^{D=3}_n,
\ee
where on the RHS $k_a^2=0$ in three dimensions implies $k_a^{\underline{\alpha} \underline{\beta}}=\lambda_a^{\underline{\alpha}}\lambda_a^{\underline{\beta}}$. One can easily verify that the procedure gives rise to the same answer regardless of the choice of $n-1$ particles.

In order to simplify notations, note that $\lambda$ and $\tilde\lambda$ transform in the same spinor representation in three dimensions and therefore we can identify $\underline{\dot\alpha}$ with $\underline{\alpha}$. 
Then using the above procedure, we are able to reduce our four dimensional amplitude given in eq.~\eqref{4dformula} to its counterpart in three dimensions ${\cal M}_{n,k}^{(s),D=3}$, which is
%
\be \int\!\! \mathbf{D}\Omega
\prod^n_{a=1}\delta^3(k_a-\oint_a dz\frac{\lambda(z)\tilde{\lambda}(z)}{P_n(z)})\delta^{0|\mathcal{N}}
(\theta_a-\oint_a
dz\frac{\lambda(z)\tilde{\eta}(z)}{P_n(z)})\delta^{0|\mathcal{N}}(\tilde\theta_a-\oint_a dz\frac{\tilde\lambda(z)\eta(z)}{P_n(z)}) I^{D=4}_s,~\label{3dpreformula}\ee
where
\be\label{reducedmeasure}
\mathbf{D}\Omega = \frac{\prod_{a=1}^nd\sigma_a}{\textrm{vol}\,(\SL2C\times \GLoneC )}\prod_{\alpha=0}^d d^{2|\frac{{\cal N}}2}\mathbb{M}_\alpha\prod_{\beta=0}^{\tilde d} d^{2|\frac{\cal N}2}\tilde{\mathbb{M}}_\beta
\prod^{n{-}1}_{b=1}\delta(\oint_b dz\frac{\l\lambda(z)
\tilde\lambda(z)\r}{P_n(z)})
\ee
and $\delta^3(p)\equiv \delta(p^{1
1})\delta(p^{2 2})\delta(p^{1 2})$.

Let us comment on some general properties before specializing to particular theories. Since $\l\lambda(z)\tilde\lambda(z)\r$ is a polynomial of
degree $d{+}\tilde d=n{-}2$, the constraints from the $\delta$ functions, i.e.~$\l\lambda(\sigma_a)\tilde\lambda(\sigma_a)\r=0$ for
$a=1,\ldots,n{-}1$, imply that the polynomial has $n-1$ distinct roots, so it must be that $\l\lambda(z)\tilde\lambda(z)\r=0$
identically. As a result of these constraints, we are able to identify
$\tilde\lambda_{\underline{\alpha}}(z)=\tau_1(z)\rho_{\underline{\alpha}}(z)$, $\lambda(z)_{\underline{\alpha}}=\tau_2(z)\rho_{\underline{\alpha}}(z)$; let us redefine the bi-degree of $(\lambda_{\underline{\alpha}}(z),\tilde\lambda_{\underline{\dot\alpha}}(z))$ as $(d',\tilde d')$, and now in three dimensions, they are given in terms of a degree-$d$ spinor polynomial $\rho_{\underline{\alpha}}(z)$, and two monic polynomials $\tau_1(z)$ and $\tau_2(z)$ with degrees $d'-d$ and $\tilde d'-d$ respectively. 

By a simple counting, we now have $n$ less variables. This is desirable, since in terms of the new polynomials which trivialize the above constraints, the original $\GLoneC$ redundancy acting on $\{\lambda_{\underline{\alpha}}(z),\tilde{\lambda}_{\underline{\dot\alpha}}(z)\}$ is partially fixed down to $\Z2$ acting on $\rho_{\underline{\alpha}}(z)$, which requires us to fix one variable, and then we need to use $n-1$ other variables to integrate out the $n-1$ constraints in \eqref{reducedmeasure}.
As we will explain in details in sec.~\ref{sec:3dcounting}, sectors in four dimensions will be redistributed into sectors in three dimensions, labeled by $d$, where physically the relevant maps are $\tau^{(n{-}2{-}2 d)}(z)\equiv \tau_1(z)\tau_2(z)$ and $\rho^{(d)}(z)$ (here the parenthesized superscript denotes the degree of the polynomial), thus in a given sector we sum over contributions from various sectors $(d',\tilde d')$ in four dimensions.



While eq.~(\ref{3dpreformula}) is the general formula for three dimensional SYM and SUGRA, in practice we are mainly interested in the sector where $n=2d+2$. In this case we have $d=d'=\tilde{d}'$, and $\tau(z)=\tau_1(z)=\tau_2(z)=1$. Due to this, the summation over $d',\tilde d'$ has only one term, and there is no need to distinguish $\theta$ and $\tilde{\theta}$. So we are allowed to combine them into the usual representation of the complete supermultiplet in three dimensions, which we also denote as $\theta$ in a slight abuse a notation. We can easily integrate out the constraints and obtain the final formula in this sector,
\be {\cal M}^{D=3}_{n=2d+2}=\int \frac{d^n\sigma\,d^{(2|\frac{\mathcal{N}}2)(d{+}1)}\mathbb{M}}{\textrm{vol}\,\SL2C\times
\Z2}
\prod^n_{a=1}\delta^3(k_a-\oint_a dz \frac{\rho(z)\rho(z)}{P_n(z)})\delta^{0|\mathcal{N}}(\theta^I_a-\oint_a dz \frac{\rho(z)\eta^I(z)}{P_n(z)})\frac{V_nI^{D=4}_s}{R(\rho)}.~\label{3dformula}\ee
Here $V_n$ is the Vandermonde determinant defined as
$V_n=\prod_{a<b}(\sigma_{a}-\sigma_b)$, and we have redefined the number of supersymmetries to be $\mathcal{N}=8s$ in order to match standard notations in three dimensions. We have again denoted the set of all coefficients by $\mathbb{M}\equiv (\rho|\chi^I,\tilde\chi^I)$; note there is a residual $\Z2$ symmetry, $\mathbb{M} \to -\mathbb{M}$, which is a remnant of the four dimensional $\GLoneC$, and corresponds to the little group in three dimensions. We have also introduced a short hand notation for the resultant $R(\rho)=R(\rho_{\underline{1}}(z),\rho_{\underline{2}}(z),z)$. This new notation is meant to remind us that these resultants are only polynomials in the coefficients of $\rho$. In the new notation note that the four dimensional integrand for gravity is simply $I^{D=4}_{s=2}=R(\tau_1\rho)R(\tau_2\rho)$.

\subsection{Maximal Supergravity}

Three-dimensional maximal super Yang-Mills amplitudes can be obtained from~(\ref{3dpreformula}) by using $I^{D=4}_{s=1}$, (\ref{parke}). Let us focus on the more interesting case of three-dimensional $\mathcal{N}=16$ supergravity; the theory was originally constructed in~\cite{Marcus:1983hb}, and on shell it can be obtained from the dimensional reduction of maximal supergravity in four dimensions. One of the most well known facts about the theory~\cite{Marcus:1983hb} is that its $n$-particle amplitudes with $n=2m+1$ vanish while those with $n=2m$ do not separate in sectors as the four dimensional theory does. Moreover, if one takes four dimensional amplitudes in a sector $k$ such that $n\neq 2k$, then after identifying $\lambda_a$ with $\tilde\lambda_a$ for all particles one gets zero.

Let us see how all these features are simple consequences of our construction, eq.~(\ref{3dpreformula}). Recall that the degree of $\tau(z)=\tau_1(z)\tau_2(z)$ is $d{-}\tilde d$; and therefore the only case when both $\tau_1(z)$ and $\tau_2(z)$ are constants is when $d=\tilde d$. Although generically two components of $\rho$ do not have common roots, in any sectors other than $d=\tilde d$, either $(\tilde\lambda_{\underline 1}(z),\tilde\lambda_{\underline 2}(z))$ or $(\lambda_{\underline 1}(z), \lambda_{\underline 2}(z))$ must
have common roots (those of $\tau_1(z)$ or $\tau_2(z)$), thus we have either $R(\tau_1\rho)=0$ or $R(\tau_2\rho)=0$. Therefore, unlike SYM, the integrand for three dimensional SUGRA and thus its amplitude vanishes unless $d=\tilde d$, which together with $d+\tilde d=n-2$ immediately implies that $n$ must be even and equal to $2k$ since $d=k-1$. Eq.~(\ref{3dformula}) with $s=2$ thus gives all non-vanishing SUGRA amplitudes in three dimensions:
\be
{\cal M}_{n=2d{+}2}^{\textrm{sugra}}=\int \frac{d^n\sigma d^{(2|8)(d{+}1)}\mathbb{M}}{\textrm{vol}\,\SL2C\times \Z2}~\prod^n_{a=1}\delta^3(k_a-\oint_a
dz\frac{\lambda(z)\lambda(z)}{P_n(z)})\delta^{0|16}(\theta_a-\oint_a dz\frac{\lambda(z)\eta(z)}{P_n(z)})V_n R(\rho).~\label{3dsugra}
\ee
Recall that $V_n = \prod_{a<b}(\sigma_a-\sigma_b)$.

\section{Scattering in Various Three-Dimensional Theories}

\subsection{Less-than-Maximal Pure SUGRA and SCS Theories}

For pure SYM and SUGRA with less supersymmetries~\footnote{In our definition, pure SYM and SUGRA means those obtained by truncating supersymmetries from the maximal cases.}, all tree
amplitudes can be extracted from eqs.~(\ref{4dformula}) and (\ref{3dpreformula}) by supersymmetry truncations. For example, it is straightforward to derive formulas with $0\leq
\mathcal{N}\leq 4s$ supersymmetries in four dimensions by integrating out (or setting to zero) components of Grassmann on-shell variables in
eq.~(\ref{4dformula}), according to the $n$ external states. However, we are mostly interested in the three dimensional cases: as we will see
shortly, the truncation of (\ref{3dsugra}) to $\mathcal{N}<16$ SUGRA theories gives a very simple and illuminating formula, which immediately
motivates a similar formula for amplitudes in supersymmetric Chern-Simons theories. 

It is well known that the S-matrix in SUGRA and that in SCS theories bear certain similarities: it is non-vanishing only for even number of
particles, with a unique R-charge sector: $k=\frac n 2$, and, for less supersymmetric cases, it is invariant only under permutations in
$S_{\frac n 2}\times S_{\frac n 2}$. With less-than-maximal supersymmetries, chiral and anti-chiral multiplets of
SU($\frac{\mathcal{N}}2$) are needed (we only consider cases when $\mathcal{N}$ is even), which can be combined into superfields $\Phi$ and $\bar\Phi$
respectively. The choice of $\Phi$ or $\bar\Phi$ for each external particle breaks the permutation invariance $S_n$, and we denote the set of
particles with $\Phi$ ($\bar \Phi$) by $E$ ($O\equiv \{1,\ldots,n\}\backslash E$).

To generalize (\ref{3dsugra}) to theories with $\mathcal{N}<16$ supersymmetries, we truncate $\mathcal{N}=16$ superfields in $E$ and $O$ to
the chiral and anti-chiral superfields $\Phi$ and $\bar\Phi$, respectively. This is achieved by integrating out $\eta^I_{a\in O}$ and setting
$\eta^I_{a\in E} \to 0$, for $I=\frac{\mathcal{N}}2{+}1,\ldots, 8$~\cite{Elvang:2011fx}. Since we work with re-defined amplitudes, we also need
to remove the wave-functions $\delta^{0|8-\frac{\mathcal{N}}2}(\lambda_a\theta^I_a)$, thus the correct procedure is to integrate over
$\theta^I_a$'s, \be \prod^8_{I=\mathcal{N}/2{+}1} \int \prod^n_{a=1}d\theta_a^I \prod_{a\in E} \eta^I_a
{\cal M}^{\textrm{sugra}}_{n,\mathcal{N}=16}={\cal M}^{\textrm{sugra}}_{n,\mathcal{N}}=\prod^n_{a=1}\delta(k_a^2) \prod^{\mathcal{N}/2}_{I=1}
\delta(\l\lambda_a\theta^I_a\r)M^{\textrm{sugra}}_{n,\mathcal{N}} (\Phi_{a\in E},\bar\Phi_{a\in O}).\ee From the RHS of (\ref{3dsugra}), the integral
removes the delta functions with $\theta_a^I$ for $I=\frac{\mathcal{N}}2{+}1,\ldots, 8$,
\be \int \prod_{a=1}^n d\theta^I_a\, \prod_{a\in E}
\eta^I_a\,\delta^{0|2}(\theta_a-\oint_a dz\frac{\lambda(z) \eta^I(z)}{P_n(z)}) =\prod_{a\in E} \eta^I(\sigma_a)\prod_{b\neq a}(\sigma_a-\sigma_b)^{-\frac 12},\ee
where the factor with square root comes from $\lambda_a=\lambda(\sigma_a)\prod_{b\neq a}\sigma_{ab}^{-\frac 12}$ and
$\lambda_a\eta_a=\frac{\lambda(\sigma_a)\eta(\sigma_a)}{\prod_{b\neq a}\sigma_{ab}}$, and we have used the abbreviation $\sigma_{ab}=\sigma_a-\sigma_b$. For each $I$ in the range, there are $|E|$ Grassmann delta
functions and $d{+}1$ integrals over the coefficients $\chi^I$, thus the result vanishes unless $|E|=d{+}1=\frac n 2$, which is nothing but the
well-known fact that $E$ and $O$ must each contain half of the external particles! For $\prod_{a\in E} \eta^I(\sigma_a)$, the Grassmann integral
over $\chi^I$ produces a Vandermonde of $\sigma$'s in $E$, \be \int d^{d{+}1}\eta^I \prod_{a\in E} \eta^I(\sigma_a)\prod_{b\neq
a}\sigma_{ab}^{-\frac 12}=\left(\prod_{a<c\in E} \sigma_{ac}\right)\left(\prod_{a\in E}\prod_{b\neq a}\sigma_{ab}^{-\frac 12}\right)
=\left(\prod_{a\in E, b\in O}\sigma_{a b}\right)^{-\frac 12}\equiv W_n^{-\frac 1 2},\ee where we have defined a new object, $W_n$, as the
product of all differences between $\sigma$'s in $E$ and those in $O$. The appearance of this factor reflects the fact that the amplitude with less
supersymmetries is only symmetric under permutations within $E$ and $O$.

Although we are mostly interested in SUGRA amplitudes, at least for the case $d=\tilde d$ it is trivial to apply the above argument to SYM
amplitudes with less-than-maximal supersymmetries. Thus we can write a unified formula for SUGRA and SYM with $d=\tilde d$,


\be {\cal M}^{D=3}_{n,(s),\mathcal{N}}=\int \frac{d^n\sigma\,d^{(2|\frac{\mathcal{N}}2)(d{+}1)}\mathbb{M}}{\textrm{vol}\,\SL2C\times
\Z2}\prod^n_{a=1}\delta^3(k_a-\oint_a dz \frac{\lambda(z)\lambda(z)}{P_n(z)})\delta^{0|\mathcal{N}}(\theta^I_a-\oint_a dz \frac{\lambda(z)\eta^I(z)}{P_n(z)})
I^{(\mathcal{N})}_s,~\label{3dlesssusy}\ee where the range of the index $I$ is $\{1,\ldots \mathcal{N}\}$, and the integrands are, \be
I^{(\mathcal{N})}_{s=1}=\frac{V_n (W_n)^{\frac{\mathcal{N}{-}8}4}}{(\sigma_{12}\sigma_{23}\cdots\sigma_{n1})R(\rho)},\quad
I^{(\mathcal{N})}_{s=2}=R(\rho)V_n (W_n)^{\frac{\mathcal{N}{-}16}4}.
\ee
Note that, apart from the Grassmann part, the only dependence of
(\ref{3dlesssusy}) on $\mathcal{N}$ is through the power of $W_n$. In~\cite{deWit:1992up} it has been argued that SUGRA theories are unique for $\mathcal{N}>8$. Here we have focused on  even $\mathcal{N}$ and we find that for $\mathcal{N}=16,12,10$, eq.~(\ref{3dlesssusy}) with
$s=2$ gives the correct tree amplitudes in the corresponding SUGRA theory. In addition, for $\mathcal{N}=8$ the formula gives tree
amplitudes in $\mathcal{N}=8$ SUGRA with 2 matter multiplets, which is the dimensional reduction of pure half-maximal SUGRA in four
dimensions.

The presence of $W_n$ in less-supersymmetric SUGRA amplitudes is very suggestive: it is a basic ingredient for three dimensional theories especially in the sector $d=\tilde d=\frac n 2-1$. The most notable example is the class of Super-Chern-Simons (SCS) theories, whose tree amplitudes exist for $d=\tilde d$. Moreover, color-ordered amplitudes in ABJM
theory are invariant under cyclic permutations in even/odd sector (BLG theory with $\mathcal{N}=8$ can be viewed as a special case of ABJM for
the gauge group $SU(2)\times SU(2)$~\cite{VanRaamsdonk:2008ft}, and the color-ordered amplitudes in BLG can be obtained from those in ABJM). Unlike three-dimensional SYM, tree amplitudes in SCS theories are conformally invariant, thus, as we discussed in sec.~\ref{sec:4dintegrands}, no resultant is needed.

Therefore, it is natural to write a formula for color-ordered tree amplitudes in SCS theories in terms of $\sigma_{12}\cdots\sigma_{n1}$ and
$W_n$. The power of these two factors can be uniquely fixed by the $\SL2C$ invariance and correct factorizations; in addition, these constraints
also determine the choice of $E$ and $O$: they must correspond to the sets of even and odd labels, which is the well-known fact that only
alternating assignment of $\Phi$ and $\bar \Phi$ is allowed for color-ordered amplitudes in SCS theories. We propose a formula for SCS amplitudes with $\mathcal{N}=6,8$:
\be {\cal M}^{\textrm{sCS}}_{n,\mathcal{N}}=\int
\frac{d^n\sigma\,d^{(2|\frac{\mathcal{N}}2)(d{+}1)}\mathbb{M}\,(W_n)^{\frac{\mathcal{N}{-}4}4}}{\textrm{vol}
\,\SL2C\times\Z2\,(\sigma_{12}\cdots\sigma_{n1})}\prod^n_{a=1}\delta^3(k_a-\oint_a dz
\frac{\lambda(z)\lambda(z)}{P_n(z)})\delta^{0|\mathcal{N}}(\theta_a-\oint_a dz
\frac{\lambda(z)\eta(z)}{P_n(z)})
,~\label{3dscs} \ee where we have a factor of $W_n^{\frac 12 }$ and $W_n$ for $\mathcal{N}=6$ and $\mathcal{N}=8$, corresponding to ABJM and BLG
theories respectively. Note that our formula for SCS amplitudes is only defined for $d=\tilde d$ sectors (SUGRA formula is defined generally, but vanishes in all other sectors), and we will return to this point shortly.

We have explicitly checked that the formula (\ref{3dscs}) agrees with known results in~\cite{Agarwal:2008pu,Bargheer:2010hn} for four and six points, and we established its validity by showing that it gives all correct factorizations. In Appendix~\ref{sec:equivalence_ABJM}, we prove the equivalence between this formula and the formula proposed in \cite{Huang:2012vt}.

\subsection{Counting Solutions in Three Dimensions}~\label{sec:3dcounting}

Here we count the solutions in three dimensions, which turns out to have a new pattern (different from that in four dimensions), and implies interesting consequences for various theories under consideration. The number of solutions does not change upon a direct dimensional reduction, so naively we have ${n{-}3 \bangle k{-}2}$ solutions for the ``sector" reduced from the four-dimensional R-charge sector $k$. However, as we discussed above, in three dimensions there are additional constraints of the form $(\lambda^{(d')}, \tilde\lambda^{(\tilde d')})=(\tau_1, \tau_2) \rho^{(d)}$, where $\tau_1 \tau_2 =\tau$ and $\rho$ are the relevant polynomials in three dimensions. This means there are $\lfloor \frac n 2{-}1\rfloor$ sectors labeled by $d$ (as opposed to $n{-}3$ in four dimensions, labeled by $(d',\tilde d')$). The equations for $n$ particles in the sector $d$ are, for $a=1,\ldots, n$,
\be \tau(\sigma_a)\rho^{\underline\alpha}(\sigma_a)\rho^{\underline\beta}(\sigma_a)=k^{\underline{\alpha \beta}}_a \prod_{b\neq a} (\sigma_a-\sigma_b). ~\label{3deq}\ee

As we have discussed, solutions in the sector $(\lambda^{(d')},\tilde\lambda^{(\tilde d')})$, reduced from sector $(d',\tilde d')$ in four dimensions with $\tilde d'\geq d'$, will fall into sectors $d=1,\ldots, d'$ in three dimensions, corresponding to splitting the $n{-}2{-}2d$ roots of $\tau(z)$ into two subsets with $d'{-}d$ and $\tilde d'{-}d$ roots, which are roots of $\tau_1(z)$ and $\tau_2(z)$ respectively. For each $d$, there are ${n{-}2{-}2d \choose d'{-}d}$ ways to split the roots, and they are indistinguishable from the three-dimensional point of view, since they correspond to the same $\rho$ and $\tau$; the sector $d$ receives indistinguishable contributions from $d'=d,d{+}1,\ldots, n{-}2{-}d$. Therefore, we have seen that each distinct solution in sector $d$ in three dimensions will have a multiplicity $m_{n,d}=\sum^{n{-}2{-}d}_{d'=d} {n{-}2{-}2d \choose d'{-}d}=2^{n{-}2{-}2d}$, which is the number of equivalent ways of producing the same $\tau$ and $\rho$ from $\lambda$ and $\tilde\lambda$; in particular, only for $d=\tilde d$ we have solutions with multiplicity one.

Denote the number of distinct solutions in sector $d$ as $N^{\textrm{dis}}_{n,d}$. We then have a splitting of the $N_{d;,\tilde d'}$ solutions into multiplets,
\be N_{d',\tilde d'}=\sum^{d'}_{d=1} {n{-}2{-}2d \choose d'{-}d} N^{\textrm{dis}}_{n,d}=\sum^{d'}_{d=1} N^{\textrm{dis}}_{n,d}\times 2^{n{-}2{-}2d}, ~\label{relation3d}\ee
and the same equality with $d', \tilde d'$ exchanged. Now we can recursively solve $N^{\textrm{dis}}_{n,d}$ from these relations, and fully determine the counting and multiplicities of solutions in three dimensions. For the $d=1$ sector, from (\ref{relation3d}) with $d'=1$ we find $N^{\textrm{dis}}_{n,1}=1$, and by denoting the number of all solutions (counting multiplicity) as $N^{D=3}_{n,d}= N^{\textrm{dis}}_d\times 2^{n{-}2d{-}2}$, this is one solution with multiplicity $2^{n{-}4}$: $N^{D=3}_{n,1}=1\times 2^{n{-}4}$.  For $d=2$, $n=6$, (\ref{relation3d}) with $d'=2$ reads $4=1 \times 2 +N^{\textrm{dis}}_{6,2}\times 1$, thus we have 2 solutions with multiplicity 1, $N^{D=3}_{6,2}=2\times 1$. It is straightforward to proceed to higher $n$ and $d$, and prove the following recursion,
\be N^{D=3}_{n,d}=2(n{-}2d{+}1) \times N^{D=3}_{n{-}1, d{-}1}\times 2^{-1}+ d \times N^{D=3}_{n{-}1,d}\times 2,\ee
where it is implied that the number of distinct solutions satisfies $N^{\textrm{dis}}_{n,d}=2(n{-}2d{+}1) N^{\textrm{dis}}_{n{-}1,d{-}1} + d N^{\textrm{dis}}_{n{-}1,d}$, and in the two terms, the multiplicity are halved and doubled respectively.  A few further examples are $N^{D=3}_{7,2}=8\times 2$, $N^{D=3}_{8,2}=22\times 4$, $N^{D=3}_{8,3}=16\times 1$.

Recall that $N_{d,\tilde d}$ is given as the Eulerian number ${n{-}3 \bangle k{-}2}$. Interestingly, the Eulerian number ${p \bangle q}$ has an interpretation as the number of permutations of $1,\ldots, p$ with $q$ ascents\footnote{For a permutation $\sigma$, $\sigma(i)$ is called an ascent (descent) if $\sigma(i)>\sigma(i{-}1) (\sigma(i)<\sigma(i{-}1))$, and it is called a peak if $\sigma(i{-}1)<\sigma(i)>\sigma(i{-}1)$.}. In three dimensions we have $N^{D=3}_{n,d}\equiv {n{-}3 \brace k{-}2}$ where the number ${ p \brace q}$ has the interpretation as the number of permutations of $1,\ldots, p$ with $q$ peaks (and $N^{\textrm{dis}}_{n,d}\equiv {n{-}3 \brace k{-}2}'$ where ${p \brace q}'$ is the number of permutations with the additional property that each ascent is immediately followed by a peak)~\cite{NewNumbers}. In particular, for the case we are mostly interested in, $n=2d{+}2$, there are $N^{D=3}_{2d{+}2,d}=N^{\textrm{dis}}_{2d{+}2,d}\times 1=E_{2d{-}1}$ roots, where $E_{2d{-}1}=E_{n{-}3}$ is the tangent number (or Euler zag number),\be \tan(x)=\sum_{p=1}^{\infty} \frac{E_{2p{-}1}}{(2p{-}1)!} x^{2p{-}1}=x+\frac {2 x^3} {3!}+\frac{16 x^5}{5!}+\frac {272 x^7}{7!}+\ldots . \ee

We have seen that for supergravity theories, any solution except for those in the $n=2d{+}2$ case make the resultant in (\ref{3dpreformula}) vanish, thus we are left with a sum over $E_{n{-}3}$ solutions in (\ref{3dsugra}). In the formula for ABJM amplitudes (\ref{3dscs}), we only sum over these $E_{n{-}3}$ solutions since there are no other sectors, and this fact indicates that it cannot come from the dimensional reduction of any four-dimensional formula.

\subsection{Infrared Behavior, Resultants and KLT Relations}

As we have discussed briefly, resultants in our formulas are connected to the contribution of zero-momentum particles in the integral over the moduli space of maps. In the computation of an actual physical scattering amplitude one integrates the amplitude for plane waves against general wave profiles over all momentum vectors $k_a^\mu$. This integration over the full space of kinematic invariants will translate into a integration over the internal space. Here we show that infrared regions will explore the tip of the null cone in the internal space and thus the power of a resultant in the integrand will determine the behavior of the theory. We study the connection in details for all the theories under consideration and show that in the integrand the resultant is the only object sensitive to soft limits.



For simplicity we focus on a single soft limit, for example when $\epsilon \to 0$ in $k^{\mu}_n=\epsilon^2 e^{\mu}_n$ where $e^{\mu}_n$ denotes the direction of the momentum of particle $n$. 
Note that in the limit, all the differences of $\sigma$ variables remain finite, so all other building blocks except the resultant: $(\sigma_{12}\cdots\sigma_{n1}), V_n, W_n$, are unaffected by the soft limit. The resultant, on the other hand, vanishes in the limit. In four dimensions, $k_n\to 0$ implies that, up to higher orders in $\epsilon^2$, either the two components of $\lambda(z)$ or those of $\tilde\lambda(z)$ develop a common root, and we find either $R(\rho)\sim \epsilon^2$ or $R(\tilde\rho)\sim \epsilon^2$; in three dimensions, since we are only interested in $d=\tilde d$ case, the soft limit implies $\lambda(\sigma_n)\sim \epsilon$ and the resultant $R(\rho)\sim \epsilon$.

Thus we have seen that in the integrand the power of resultant is directly related to the power of $\epsilon$ in the soft limit; 
as $k_n=\epsilon^2 e_n \to 0$, we find
\be I^{\textrm{sYM}}_{n,D=4}\sim \epsilon^0,\, I^{\textrm{sugra}}_{n,D=4}\sim \epsilon^2,\quad I^{\textrm{sYM}}_{n=2d{+}2,D=3}\sim \epsilon^{-1},\, I^{\textrm{sCS}}_n\sim \epsilon^0,I^{\textrm{sugra}}_{n,D=3}\sim \epsilon,\ee
for any amount of supersymmetries under consideration. This clearly demonstrates that the integrand for gravity has better infrared behavior than that of Yang-Mills, and in three dimensions, Chern-Simons is also better behaved than Yang-Mills. By considering the maps and integral measure, it is easy to see that for maximal supergravity in four and three dimensions, the single soft scalar emission always vanishes, which reflects the existence of a moduli space; it would be interesting to study double soft scalar emissions in order to reveal the underlying $E_{7(7)}$ or $E_{8 (8)}$ symmetries respectively.

Just as Yang-Mills in four dimensions, conformally invariant Chern-Simons in three dimensions cannot distinguish zero momentum from any null momentum, which explains why no resultant appears in the formula; the pathological infrared behavior of three-dimensional Yang-Mills may be traced back to the presence of a resultant in the denominator of its integrand.

It is well known that tree amplitudes in gravity and Yang-Mills are related to each other by the Kawai-Lewellen-Tye (KLT) relations. Since our formula (\ref{3dformula}) and (\ref{3dlesssusy}) are reduced from four dimensions, they are guaranteed to satisfy KLT relations: $8\leq \mathcal{N}\leq 16$ supergravity tree amplitudes are given by the KLT bilinear form of tree amplitudes in maximal $\mathcal{N}_L=8$ SYM and in $\mathcal{N}_R=\mathcal{N}-8$ SYM. It is not obvious at all, however, how the relations work for our formula in three dimensions, especially why the bad infrared behavior of Yang-Mills does not propagate to gravity. 

Our derivation of KLT in three dimensions parallels that in~\cite{Cachazo:2012da} in four dimensions, and here we only give the result. A remarkable fact, which has been proved and will be presented in~\cite{Cachazo2013}, is the orthogonality of the solutions with respect to the KLT bilinear form acting on the vector space of partial amplitudes, \emph{i.e.} $\sum_{\alpha,\beta} \frac 1 {\alpha(\sigma_{12}\sigma_{23}\cdots\sigma_{n1})^{(i)}}S[\alpha|\beta] \frac 1 {\beta(\sigma_{12}\sigma_{23}\cdots\sigma_{n1})^{(j)}}=0$ for any two different solutions $i\neq j$.

In three dimensions, this means gravity amplitudes (which are non-vanishing only for $n=2d{+}2$) obtained from KLT relations involve a sum over $E_{n{-}3}$ (as opposed to $E^2_{n{-}3}$) solutions, which happen to be the only solutions with multiplicity one. Following an argument similar to that in~\cite{Cachazo:2012da}\footnote{In particular, it is crucial to use the fact $\sum_{b\neq a} \l \lambda_a \lambda_b\r^2/\sigma_{ab}=0$ which implies the Bern-Carrasco-Johansson (BCJ) relations~\cite{Bern:2008qj} in three dimensions.}, we can show that each solution $i$ in the sum gives,
\be \sum_{\alpha,\beta\in S_{n{-}3}} \frac 1 {\alpha(\sigma_{12}\sigma_{23}\cdots\sigma_{n1})^{(i)}}S[\alpha|\beta] \frac 1 {\beta(\sigma_{12}\sigma_{23}\cdots\sigma_{n1})^{(i)}}=\frac{\det H^{(i)}_n}{J^{(i)}_n{}^2}, \ee
where on the RHS we have the Jacobian of all bosonic delta functions in (\ref{4dformula}), $J_n$, and a determinant, $\det H_n(\rho)$ constructed from the map $\lambda(z)$. The definition and examples of $H_n$ can be found in Appendix~\ref{sec:resultant}. A new feature in three dimensions is that for $n>4$, $\det H_n(\rho)=m(\rho) R^3(\rho)$ with a polynomial $m(\rho)$, and $J_n=V_n m(\rho)$. By using one power of the Jacobian we can rewrite the sum as a integral over rational maps, and reproduce (\ref{3dsugra}) and (\ref{3dlesssusy}) for three-dimensional SUGRA; it is easy to see that the integrand is \be
\frac{\det H_n(\rho)}{J_n}\frac{V_n^2}{R(\rho)^2}(W_n)^{\frac{\mathcal{N}_R-8}4}=V_n R(\rho)(W_n)^{\frac{\mathcal{N}-16}4}.\ee Thus we see that it is exactly the factor $R^3$ in $\det H_n$, coming from KLT relations, that turns the integrand from $R^{-2}$ to $R$, and produces a good infrared behavior.

\section{Conclusions and Future Directions}

By dressing the S-matrix with proper wave functions and exploiting the idea of rational maps from $\CP1$ to the null cone in momentum space, we derived formulas for tree-level scattering amplitudes in SYM, SUGRA and SCS theories in three dimensions. The rational maps, eq.~(\ref{4dformula}) and its three-dimensional reduction, provide a universal description for tree-level scattering processes in all these theories, and the spinor-helicity variables make four and three dimensions particularly simple. Tree amplitudes are given by summing over solutions from solving (\ref{4deq}) or (\ref{3deq}) which are independent of theories (except for SCS cases one has to select a special subset), with theory-dependent integrands evaluated at those solutions. We have understood the counting and structures of the solutions, and it would be very interesting to further study properties and possible combinatoric interpretations of them.

The integrands in all these cases are given by simple monomials of four basic objects, $\sigma_{12}\cdots \sigma_{n1}, R, V_n, W_n$, where the first one is only present for theories with color orderings, and the last one only in three dimensions (in particular it is an essential input for SCS formulas). We have discussed the connections between powers of resultants, infrared behavior, zero-momentum configurations and conformal properties in all these theories, and it is worth further exploring these connections and implications for the moduli space of supergravity. In addition, we have discussed KLT relations in three dimensions, and it would be interesting to see if our formalism can shed some lights on related double-copy relations, especially the ones between three-dimensional SCS and SUGRA theories, which are not completely understood~\cite{Bargheer:2012gv,Huang:2012wr}.

\subsection{General $\SL2C$-Invariant Formula}

An important open question is whether our construction can be extended to other physically sensible scattering amplitudes in four and three dimensions. Since (\ref{4dformula}) is universal, the key is to constrain other possible integrands using $\SL2C$ invariance and physical factorization properties. Under the assumption of using only these four building blocks, one can fully classify the space of integrands which respect $\SL2C$ invariance in three dimensions. In addition to eqs.~(\ref{3dformula}), (\ref{3dlesssusy}), (\ref{3dscs}), there exists a large zoo of $\SL2C$-invariant formulas. Restricting to $d=\tilde d$ case, we have a general $\SL2C$-invariant formula for ${\cal M}^{D=3}_{n=2d{+}2}$,
\be \int \frac{d^n\sigma d^{(2|\frac{\mathcal{N}}2)(d{+}1)}\mathbb{M}}{\textrm{vol}\,\SL2C\times \Z2}~\prod^n_{a=1}\delta^3(k_a-\oint_a dz
\frac{\lambda(z)\lambda(z)}{P_n(z)})\delta^{0|\mathcal{N}}(\theta_a-\oint_a dz \frac{\lambda(z)\eta(z)}{P_n(z)})\frac {V_n^v W_n^{\frac{\mathcal{N}}4-w} R(\rho)^{w-2v-1}}{(\sigma_{12}\cdots \sigma_{n1})^{\frac{3+v-w}2}},~\label{general3d}\ee
which depends on three parameters $\mathcal{N},v,w$. To give physically sensible amplitudes, the formula should only have physical poles and correct factorizations on them, which encode locality and unitarity. Formulas for SUGRA, SYM and SCS, which correspond to $(v,w)=(1,4),(1,2),(0,1)$ respectively, all satisfy the constraints; we have also performed a partial analysis on  factorization properties for other classes of amplitudes in~(\ref{general3d}). It would be fascinating to completely determine the classes of formulas which give sensible amplitudes and their corresponding theories.


\subsection{Scattering Equations}

The starting point in this work was the study of four dimensional amplitudes dressed with wave functions. Amplitudes were written in (\ref{4dformula}) as an integral over the moduli space of certain rational maps. One crucial ingredient is that the kinematic space given by $n$ momentum vectors $k_a^\mu$ is connected to the internal space via
\be
k^\mu_a = \frac{1}{2\pi i}\oint_{|z-\sigma_a|=\epsilon} dz \frac{k^\mu(z)}{\prod_{b=1}^n(z-\sigma_b)}
\ee
with $k^\mu(z)$ a vector of degree $n-2$ polynomials satisfying $k^\mu(z)k_\mu(z) =0$. Note that in these equations no reference is made to four dimensions and, in fact, this is implicitly what allowed us to go straightforwardly down to three dimensions. It is obviously tempting to take these equations as a fundamental set of equations for scattering of massless particles in any number of dimensions. Note that the requirement that $k^{\mu}(z)$ be a null vector for any $z$ means that the whole $\mathbb{CP}^1$ is mapped to the null cone in momentum space. This is much stronger than simply requiring that $k(\sigma_a)^2=0$ for all $n$ values of $a$. To see this note that $k(z)^2$ is a polynomial of degree $2n-4$ and knowing $n$ roots of it is not enough to find it. We need an additional $n-3$ conditions -- $k(z)^2$ is not monic. It is a simple exercise to show that the new $n-3$ conditions can be written as
\be
k(\sigma_a)\cdot k'(\sigma_a)=0
\ee
for $a$ taking values in any subset of $n-3$ elements of $\{1,\ldots ,n\}$ and where $k'_\mu (z)$ is the derivative of the polynomials $k_\mu(z)$ with respect to $z$.

These equations can also be written in a more suggestive form by using that $k^\mu(\sigma_a) = k^\mu_a \prod_{b\neq a}(\sigma_a-\sigma_b)$ to find
\be
k(\sigma_a)\cdot k'(\sigma_a)=v_a^2 \sum_{b\neq a} \frac{s_{ab}}{\sigma_{a}-\sigma_b}
\ee
with $\displaystyle v_a\equiv \prod_{b\neq a} (\sigma_{a}-\sigma_b)$. Using that $v_a$ can never vanish for generic momenta we find that the equations become
\be
\sum_{b\neq a} \frac{s_{ab}}{\sigma_{a}-\sigma_b} = 0 \quad {\rm with} \quad a\in\{1,\ldots ,n\}.~\label{scattering}
\ee
We would like to call these the {\it scattering equations}. Of course, only $n-3$ of these equations are linearly independent. In~\cite{Cachazo2013}, we explore these equations and prove some of their remarkable properties.

Finally, it is important to note that these equations were first encountered in \cite{Cachazo:2012uq} in the form
\be
\sum_{b\neq a} \frac{s_{ab}(\sigma_c-\sigma_b)}{\sigma_{a}-\sigma_b} = 0
\ee
where $c$ is some fixed label. Writing the numerator of each term as $(\sigma_c-\sigma_b) = (\sigma_c-\sigma_a) + (\sigma_a - \sigma_b)$ and using momentum conservation, $\sum_{b}s_{ab} = 0$, one gets (\ref{scattering}). In \cite{Cachazo:2012uq} the scattering equations were shown to be at the heart of the fundamental BCJ identity~\cite{Bern:2008qj} satisfied by Yang-Mills amplitudes. The fact that the BCJ relation is a property of Yang-Mills amplitudes in any dimension is yet another hint that the scattering equations can play a fundamental role in arbitrary dimensions.

\acknowledgments

This work is supported by the Perimeter Institute for Theoretical Physics. Research at Perimeter Institute is supported by the Government of Canada through Industry Canada and by the Province of Ontario through the Ministry of Research \& Innovation.

\appendix

\section{Basic Facts on Resultants}\label{sec:resultant}

In the discussion on the integrand $I_s$, we have seen that the resultant of the two components of the polynomial maps $\lambda_{\underline{\alpha}}(z)$ or $\tilde{\lambda}_{\underline{\dot{\alpha}}}(z)$ with respect to $z$ may naturally arise when considering the proper target space for graviton amplitudes.

In general, if given any two polynomials $F(z)$ and $G(z)$, with degree $d_f$ and $d_g$ respectively, and we express them in terms of their roots
\begin{equation}
F(z)=F_{d_f}\prod_{m=1}^{d_f}(z-a_m),\quad\quad G(z)=G_{d_g}\prod_{m=1}^{d_g}(z-b_m),
\end{equation}
then the resultant of these two polynomials with respect to their common variable $z$ can be defined as
\begin{equation}
R(F(z),G(z),z)=F_{d_f}^{d_g}G_{d_g}^{d_f}\prod_{i=1}^{d_f}\prod_{j=1}^{d_g}(a_i-b_j).
\end{equation}
This means that the resultant is the criteria of checking whether the two polynomials share at least one common root.

Without loss of generality, let's assume that $d_g\leq d_f$, then the resultant can be alternatively computed by first constructing a $d_g\times d_g$ matrix $\|R\|$, whose entries are obtained by
\begin{equation}\label{eq:determinantmatrix}
\|R\|_{i,j}=\oint_{\infty}\frac{dx}{x^{i}}\oint_{\infty}\frac{dy}{y^{j}}\frac{F(x)G(y)-F(y)G(x)}{x-y},
\end{equation}
with $i,j\in\{1,\ldots,d_g\}$. Then we have the relation
\begin{equation}
R(F(z),G(z),z)=G_{d_g}^{d_f-d_g}\det\left(\|R\|\right).
\end{equation}

In particular, when the degrees of the two polynomials match, as the two component polynomials of $\lambda_{\underline{\alpha}}(z)$ or $\tilde{\lambda}_{\underline{\dot{\alpha}}}(z)$ in our case, the resultant is just the determinant of (\ref{eq:determinantmatrix}) itself. Furthermore, e.g.~when we just focus on $\lambda_{\underline{\alpha}}(z)$, we can observe by eq.~(\ref{eq:determinantmatrix}) that the matrix $\|R\|$ is a Lorentz--invariant quantity, and thus can be expressed solely in terms of the Lorentz--invariant spinor products $\langle\rho_{\alpha}\rho_{\beta}\rangle$ formed by the coefficients of $\lambda_{\underline{\alpha}}(z)$. In this case, the general structure of the matrix $\|R\|_{ij}$ is
\begin{equation}
\|R\|_{i,j}=\sum_{k=0}^{\min(i-1,j-1,d-i,d-j)}\langle\rho_{i-1-k}\rho_{j+k}\rangle.
\end{equation}
Similar formulas apply for $\tilde{\lambda}_{\underline{\dot{\alpha}}}(z)$. Here we list some simple examples. When $d=2$, we have
\begin{equation}
\|R\|=\left(\begin{array}{cc}
\langle\rho_{0}\rho_{1}\rangle&\langle\rho_{0}\rho_{2}\rangle\\
\langle\rho_{0}\rho_{2}\rangle&\langle\rho_{1}\rho_{2}\rangle
\end{array}\right).
\end{equation}
When $d=3$, we have
\begin{equation}
\|R\|=\left(\begin{array}{ccc}
\langle\rho_{0}\rho_{1}\rangle&\langle\rho_{0}\rho_{2}\rangle&\langle\rho_{0}\rho_{3}\rangle\\
\langle\rho_{0}\rho_{2}\rangle&\langle\rho_{1}\rho_{2}\rangle+\langle\rho_{0}\rho_{3}\rangle&\langle\rho_{1}\rho_{3}\rangle\\
\langle\rho_{0}\rho_{3}\rangle&\langle\rho_{1}\rho_{3}\rangle&\langle\rho_{2}\rho_{3}\rangle
\end{array}\right).
\end{equation}
When $d=4$, we have
\begin{equation}
\|R\|=\left(\begin{array}{cccc}
\langle\rho_{0}\rho_{1}\rangle&\langle\rho_{0}\rho_{2}\rangle&\langle\rho_{0}\rho_{3}\rangle&\langle\rho_{0}\rho_{4}\rangle\\
\langle\rho_{0}\rho_{2}\rangle&\langle\rho_{1}\rho_{2}\rangle+\langle\rho_{0}\rho_{3}\rangle&\langle\rho_{1}\rho_{3}\rangle+\langle\rho_{0}\rho_{4}\rangle&\langle\rho_{1}\rho_{4}\rangle\\
\langle\rho_{0}\rho_{3}\rangle&\langle\rho_{1}\rho_{3}\rangle+\langle\rho_{0}\rho_{4}\rangle&\langle\rho_{2}\rho_{3}\rangle+\langle\rho_{1}\rho_{4}\rangle&\langle\rho_{2}\rho_{4}\rangle\\
\langle\rho_{0}\rho_{4}\rangle&\langle\rho_{1}\rho_{4}\rangle&\langle\rho_{2}\rho_{4}\rangle&\langle\rho_{3}\rho_{4}\rangle
\end{array}\right).
\end{equation}

In the discussion of KLT relations we also see the appearance of the determinant of a matrix $H_n$, whose dimension is $(n-3)\times(n-3)$, and whose entries only explicitly depend on $\rho$'s and $\tilde\rho$'s when restricted to $3$ or $4$ dimensions. In any dimensions, $H_n$ can be defined with the most generic form of the degree-$(n-2)$ momentum polynomial $k^\mu(z)$ as appearing in \eqref{4dformula}, via the following relation
\be
(H_n)_{ij}=\oint_\infty\frac{dx}{x^i}\oint_\infty\frac{dy}{y^j}\frac{k(x)\cdot k(y)}{(x-y)^2},
\ee
where ``$\cdot$'' denotes the contraction of Lorentz indices. When restricted to $4$ dimensions, where $k_\mu(z)\sigma^\mu_{\underline{\alpha{\dot\alpha}}}=\lambda_{\underline\alpha}(z)\tilde{\lambda}_{\underline{\dot\alpha}}(z)$, since $k(x)\cdot k(y)=\langle \lambda(x)\lambda(y)\rangle[\tilde{\lambda}(x)\tilde{\lambda}(y)]$, we see that $H_n$ matrix is actually the convolution of the resultant matrices $\|R(\rho)\|$ of polynomials $\lambda_{\underline{\alpha}}(z)$ and $\|R(\tilde{\rho})\|$ of $\tilde{\lambda}_{\underline{\dot\alpha}}(z)$ as discussed above
\be
(H_n)_{ij}=\oint_\infty\frac{dx}{x^i}\oint_\infty\frac{dy}{y^j}
\left\lgroup\sum_{k,l=0}^{d-1}\|R(\rho)\|x^ky^l\right\rgroup
\left\lgroup\sum_{r,s=0}^{\tilde{d}-1}\|R(\tilde{\rho})\|x^ry^s\right\rgroup.
\ee
Furthermore, in four dimensions the determinant of $H_n$ always factorizes into three parts (specializing in this case we change the notation to $H_{d,\tilde{d}}^{\text(4d)}$)
\be\label{H4d}
\det{H_{d,\tilde{d}}^{\text(4d)}}=R(\rho)R(\tilde{\rho})M_{d,\tilde{d}}(\rho,\tilde{\rho}),
\ee
where $R(\rho)$ and $R(\tilde{\rho})$ are exactly the resultants of the two types of spinor polynomial maps respectively, and the remaining factor $M_{d,\tilde{d}}$ depends on both $\rho$'s and $\tilde{\rho}$'s in general.

When restricted to three dimensions, we can directly dimensional reduce the relation \eqref{H4d} down to the $d=\tilde{d}$ sector, where $R(\tilde{\rho})$ is now identified with $R(\rho)$. Remarkably, it turns out that the original $M_{d,\tilde{d}}$ factor produces one more $R(\rho)$ in the reduction, so that the determinant of $H_n$ (which we now denote as $H_d^{\text{(3d)}}$) can be expressed as
\be
\det{H_d^{\text{(3d)}}}=R^3(\rho)m_d(\rho),
\ee
for $n>4$. This new form of $H$ is guaranteed by three dimensional KLT relations. It has also been checked explicitly up to $d=3$ (i.e.~$n=8$). And in these cases the expression of $m_d(\rho)$ is
\be
m_2=2,\quad\quad m_3=4(\langle\rho_1\rho_2\rangle-3\langle\rho_0\rho_3\rangle).
\ee

\section{Equivalence of Two Formulas for ABJM Amplitudes}\label{sec:equivalence_ABJM}

Here we show directly that the formula (\ref{3dscs}) for $\mathcal{N}=6$ is equivalent to the recently proposed formula of Huang and
Lee~\cite{Huang:2012vt},
\be M_n=\frac 1{\textrm{vol}\,\GLtwoC}\int d^{2n}\sigma\frac{J\Delta\prod_{\alpha=1}^k\delta^{2|3}(C_{\alpha
i}(\sigma)\lambda_i)}{(12)(23)\cdots(n1)}, \ee where $\sigma_i=(a_i,b_i)$, $(i,j)\equiv a_i b_j-a_j b_i$; the factors are
$\Delta=\prod_{j=1}^{2d{+}1}\delta\left(\sum_{i}{a_i^{2d{+}1{-}j}b_i^{j-1}}\right)$, \be J=\frac{\prod_{1\leq i<j\leq2d+1}(ij)}{\prod_{1\leq
i<j\leq d{+}1}(2i{-}1,2j{-}1)}, \quad C_{\alpha
i}=\left(\begin{array}{cccc}a_1^d&a_2^d&\cdots&a_n^d\\a_1^{d-1}b_1&a_2^{d-1}b_2&\cdots&a_n^{d-1}b_n\\\vdots&\vdots&&\vdots\\a_1b_1^{d-1}&a_2b_2^{d-1}&\cdots&a_nb_n^{d-1}\\b_1^d&b_2^d&\cdots&b_n^d\end{array}\right).
\ee

First we change the variables to inhomogeneous coordinates, $(a_i,b_i)=t_i^{\frac 1 d} (1,\sigma_i)$, then the entries of the matrix become
$C_{\alpha i}=t_i \sigma^\alpha_i$, $\Delta=\prod_{\alpha=0}^{2d}\delta\left(\sum_{i}t_i^2\sigma_i^\alpha\right)$, and we have \be
\frac{d^{2n}\sigma\,
J}{(12)(23)\cdots(n1)}=\prod_{i=1}^{n}\frac{dt_i}{t_i}d\sigma_i\frac{(-1)^{\frac{d(d-1)}{2}}\prod_{i=1}^{n-1}t_i^2\prod_{1\leq i<j\leq
2d{+}1}\sigma_{i,j}}{d \prod^n_{i=1}\sigma_{i,i{+}1}\prod_{i=0}^{d}t_{2i+1}\prod_{1\leq i<j\leq d{+}1}\sigma_{2i{-}1,2j{-}1}}. \label{J}\ee The
bosonic and fermionic delta functions (written as fermionic Fourier transforms) become \ba
\prod_{\alpha=0}^{d}\delta^2\left(\sum_{i}a_i^{d-\alpha}b_i^\alpha\lambda_i\right)
&=&V_n^{-2}\int{\prod_{\alpha=0}^{d}d^2\rho_\alpha}\prod_{i=1}^{n}\frac{\lambda_i^1\lambda_i^1}{t_i^2}\delta^2(k_i-t_i^2\lambda(\sigma_i)\lambda(\sigma_i)),\nl
\int{\prod_{\alpha=0}^{d}d^{0|3}\chi_\alpha}\prod_{i=1}^{n}\delta^{0|3}(\eta_i-t_i\chi(\sigma_i))
&=&\int{\prod_{\alpha=0}^{d}d^{0|3}\chi_\alpha}\prod_{i=1}^{n}\frac{1}{\lambda_i^1}\delta^{0|3}(\theta_i^{\underline
1}-t_i^2\chi(\sigma_i)\lambda^{\underline 1}(\sigma_i)), \ea where on the RHS of the first equation, $\delta^2(p)\equiv
\delta(p^{\underline{11}})\delta(p^{\underline{22}})$. We can see that it is important to insert back the wave functions,
$\delta(k_a^2)\delta^{0|3}(\l\lambda_a\theta_a\r)$, then the delta functions can be rewritten as manifestly Lorentz invariant.

To proceed, note that using the delta functions in $\Delta$, one can solve all the $t$'s except $t_n$, which gives $t_i=\pm t_n
\left(\frac{\prod_{j\neq n}\sigma_{n j}}{\prod_{j\neq i} \sigma_{i j}}\right)^{\frac 1 2}$.  Using the $\GLoneC$ symmetry to fix $t_n$, and note
that the only place $t_i$, rather than $t_i^2$, appear is on the RHS of eq.~(\ref{J}). It is easy to see that such $t_i$ factors indeed produce
$W_n^{\frac 12}$ as we expect. In the end we arrive at eq.~(\ref{3dscs}), \be {\cal M}_n
=\int\frac{d^{n}\sigma\,d^{(2|3)(d{+}1)}\mathbb{M}\,W_n^{\frac 1 2}} {\textrm{vol}\,\SL2C\times \Z2\,\prod^n_{i=1}\sigma_{i,i{+}1}}
\prod_{i=1}^{n}\delta^3(k_i-\frac{\lambda(\sigma_i)\lambda(\sigma_i)}{\prod_{j\neq i} \sigma_{i
j}})\delta^{0|6}(\theta_i-\frac{\lambda(\sigma_i)\chi(\sigma_i)}{\prod_{j\neq i} \sigma_{i j}}).~\label{ABJM} \ee

\bibliographystyle{JHEP}
\bibliography{RationalMaps3d}

\end{document}